# Numerical Determination of the Optimal Value of Quantizer's Segment Threshold Using Quadratic Spline Functions

**Lazar Velimirović[1*], Zoran Perić[2], Miomir Stanković[3], Jelena Nikolić[2]**

[1]*Mathematical Institute of the Serbian Academy of Sciences and Arts, Belgrade, Serbia*
[2]*Department of Telecommunications, Faculty of Electronic Engineering, University of Niš, Serbia*
[3]*Mathematical Department, Faculty of Occupational Safety, University of Niš, Serbia*

.

## ABSTRACT

In this paper, an approximation of the optimal compressor function using the quadratic spline functions has been presented. The coefficients of the quadratic spline functions are determined by minimizing the mean-square error (MSE). Based on the obtained approximative quadratic spline functions, the design for companding quantizer for Gaussian source is done. The support region of proposed companding quantizer is divided on segments of unequal size, where the optimal value of segment threshold is numerically determined depending on maximal value of the signal to quantization noise ratio (SQNR). It is shown that by the companding quantizer proposed in this paper, the SQNR that is very close to SQNR of nonlinear optimal companding quantizer is achieved.

## 1. INTRODUCTION

Quantization is the nonlinear, zero-memory operation of converting a continuous signal into a discrete signal that assumes only a finite number of levels The input to a quantizer is the original data, and the output is always one among a finite number of levels [1,2]. The quantizer is a function whose set of output values are discrete, and usually finite. Obviously, this is a process of approximation, and a good quantizer is one which represents the original signal with minimum loss or distortion.

In scalar quantization, each input symbol is treated separately in producing the output. A quantizer can be specified by its input partitions and output levels (also called reproduction levels). If the input range is divided into levels of equal spacing, then the quantizer is termed as a uniform quantizer, and if not, it is termed as a nonuniform quantizer [1,2]. Uniform quantizers are suitable for signals that have approximately uniform distribution. How most of the signals do not have a uniform distribution there is a need for using nonuniform quantizer.

Nonuniform quantization can be realized through the process of companding, namely by applying an optimal compressor function on input signal. By knowing optimal compressor function, model quantizer is completely defined. However, it is shown that this model is very difficult to realize [1,2]. Also, it is well known that designing nonlinear optimal companding quantizers for a Gaussian source is very complex due to the difficulties in determining the inverse optimal compressor function [1-6]. Due to the complex process of optimal compressor functions practical realizations, their approximation is often performed.

*Corresponding author: E-mail address: velimirovic.lazar@gmail.com

In this paper we will propose a new method of construction companding quantizer which is based on the approximation of the optimal compressor function using quadratic spline functions with the support region divided into $2L = 4$ segments. The design is performed for this number of segments so that the proposed companding quantizer would have low complexity. The goal is to achieve as high as possible the signal to quantization noise ratio SQNR for as small as possible complexity [2]. In order to achieve with the proposed companding quantizer higher SQNR close to SQNR value of the nonlinear optimal companding quantizer, in this paper we propose that the coefficients based on which the approximative quadratic spline functions are formed, are determined by minimization mean square error (MSE). Based on the obtained approximative quadratic spline functions, quantizer designing process will be performed for the Gaussian source.

Support region of the quantizer considered in [4] is divided into equal segments while the number of reproduction levels inside segments is determined by optimizing the granular distortion and under the constraint of total number of reproduction levels. Unlike with the quantizer described in [4], the support region of proposed companding quantizer is not divided on segments of equal size. In this paper, the optimal value of segment threshold of proposed companding quantizer is numerically determined depending on maximal value of SQNR. The number of cells per segments is determined depending on value of optimized segment threshold. The value of segment threshold depends on the size of the support region [5].

The remainder of the paper is organized as follows: In section 2 the detailed description of the quadratic spline functions. Also, in section 2, the optimal compressor function is given. Design of companding quantizer based on approximate quadratic spline functions is described in section 3. Procedure of numerical determination of the optimal value of the segment threshold is described in section 4. Finally, section 5 presents a discussion of numerical results obtained using the proposed companding quantizer for the Gaussian source of unit variance.

## 2. APPROXIMATE QUADRATIC SPLINE FUNCTIONS

A spline curve is a mathematical representation for which it is easy to build an interface that will allow a user to design and control the shape of complex curves and surfaces. The general approach is that the user enters a sequence of points, and a curve is constructed whose shape closely follows this sequence [7]. A spline function consists of polynomial pieces on subintervals joined together with certain continuity conditions. Formally, suppose that $n+1$ points $x_0, x_1, \ldots, x_n$ have been specified and satisfy $x_0 < x_1 < \cdots < x_n$. These points are called knots. Suppose also that an integer $k \geq 0$ has been prescribed. A spline function of degree $k$ having knots $x_0, x_1, \ldots, x_n$ is a function $S$ such that [7]:

1. On each interval $[x_{i-1}, x_i]$, $S$ is a polynomial of degree $\leq k$.

2. $S$ has a continuous $(k - 1)$st derivative on $[x_0, x_n]$.

Hence, $S$ is a piecewise polynomial of degree at most $k$ having continuous derivatives of all orders up to $k$ - 1.

In this paper, the approximation of the optimal compressor function using quadratic spline functions $(k = 2)$ is done. The quadratic spline is a continuously differentiable piecewise quadratic function, where quadratic includes all linear combinations of the basic functions $x \rightarrow 1$, $x$, $x^2$. It is consisted of parabola parts between two consecutive knotes, but elected to have the same tangent at knote [7]. The approximate quadratic spline function $h(x)$, which approximates a nonlinear optimal compressed function $c(x)$, has the following form [7]:

$$h(x) = \begin{cases} r_1 + p_1 x + q_1 x^2, & x \in [0, x_1] \\ r_2 + p_2 x + q_2 x^2, & x \in [x_1, x_2] \\ \vdots & \\ r_L + p_L x + q_L x^2, & x \in [x_{L-1}, x_L] \end{cases}. \tag{1}$$

The coefficients of the quadratic spline functions, $r_i$, $p_i$ and $q_i$, $i = 1,...,L$, are determined by minimizing the mean-square error (MSE) as follows:

$$F(x) = \sum_{i=1}^{L} \frac{1}{x_i - x_{i-1}} \int_{x_{i-1}}^{x_i} (c(x) - h(x))^2 \, dx \, , \tag{2}$$

$$\frac{\partial F(x)}{\partial r_i} = 0, \; \frac{\partial F(x)}{\partial p_i} = 0, \; \frac{\partial F(x)}{\partial q_i} = 0, \; i = 1, \ldots, L \, . \tag{3}$$

The optimal compressor function $c(x)$ by which the maximum SQNR is achieved for the reference variance of an input signal is defined as [1,2]:

$$c(x) = \left\{ x_{\max} \operatorname{sgn} x \frac{\int_0^{|x|} p^{1/3}(t)dt}{\int_0^{x_{\max}} p^{1/3}(t)dt}, \quad |x| \le x_{\max} \right. , \tag{4}$$

where $x_{\max}$ denotes the support region threshold of the optimal companding quantizer and $p(t)$ is a symmetric Gaussian probability density function (PDF). Without diminishing the generality, the companding quantizer design will be done for the reference input variance of $\sigma_{ref}^2 = 1$ and $0 \le x \le x_{\max}$. In the rest of the paper we assume symmetry about zero in the companding quantizer design. The Gaussian PDF, we consider here, is indeed symmetrical about zero. The PDF of this source is given by [1,2]:

$$p(x) = \frac{1}{\sqrt{2\pi}} \exp\left( -\frac{x^2}{2} \right). \tag{5}$$

## 3. DESIGN OF COMPANDING QUANTIZER BASED ON APPROXIMATE QUADRATIC SPLINE FUNCTIONS

Two finite sets of real numbers, characterize each scalar quantizer: set of reproduction levels $\{y_1, ..., y_{max}\}$ and set of decision thresholds $\{x_0, x_1, ..., x_{max}\}$. Decision thresholds divide input range of the quantizer into $N$ quantization cells also called quantization intervals $a_j = [x_j - 1, x_j)$, $j = 1, 2, ..., N$. The support region of scalar quantizer $[-x_{max}, x_{max}]$ is divided into $L$ segments in both quadrants, where each segment is divided into specific number of cells whose size differs from segment to segment. The support region threshold of our companding quantizer is defined as follows [5]:

$$x_{max} = \hat{\sigma}\sqrt{6\ln N}\left[1 - \frac{\ln\ln N}{4\ln N} - \frac{\ln\left(3\sqrt{\pi}\right)}{2\ln N}\right]. \quad (6)$$

Let us assume, as in [4], that in the granular region $[-x_{max}, x_{max}]$, the total number of the reproduction levels per segments in the first quadrant is:

$$\sum_{i=1}^{L} \frac{N_i}{2} = \frac{N-2}{2}, \quad (7)$$

where the number of reproduction levels per segments, $N_i/2$, is determined from the following condition:

$$\frac{N_i}{2} = \frac{N-2}{2}\frac{c_i(x_i) - c_{i-1}(x_{i-1})}{c_L(x_L)}, \; i = 1, \dots, L. \quad (8)$$

Reproduction levels of the companding quantizer we propose are determined as the solution of the approximate quadratic spline functions as follows:

$$y_{i,j} = h_i^{-1}\left(\left(\frac{2j-1}{2}\right)\Delta\right), \; i = 1, \; j = 1, \dots, \frac{N_i}{2}, \quad (9)$$

$$y_{i,j} = h_i^{-1}\left(c_i(x_i) + \left(\frac{2j-1}{2}\right)\Delta\right), \; i = 2, \dots, L, \; j = 1, \dots, \frac{N_i}{2}, \quad (10)$$

where for the $y_{i,j}$ is taken the solution that belongs to the spline function domain. Indexes *i* and *j* indicate the *j*-th reproduction levels within the *i*-th segment. Step size Δ is determined with:

$$\Delta = \frac{2x_{max}}{N-2}. \quad (11)$$

The cells lengths per segments of the considered companding quantizer are determined by:

$$\Delta_{i,j} = \frac{\Delta}{h_i^{'}\left(y_{i,j}\right)}, \; i = 1, \dots, L, \; j = 1, \dots, \frac{N_i}{2}. \quad (11)$$

where by $\Delta_{i,j}$ denotes the $j$-th cells lengths within the $i$-th segment. The total distortion is a quality measure of quantization process, and can be found as a sum of the granular $D_g$ and the overload $D_o$ distortion. The granular distortion for such a designed model is [1,2]:

$$D_g = 2\frac{{x_{\max}}^2}{3(N-2)^2}\sum_{i=1}^{L}\sum_{j=1}^{N_i/2}\frac{p(y_{i,j})}{\left[h_i^{'}(y_{i,j})\right]^2}\Delta_{i,j}\ ,\ i=1,\ldots,L,\ j=1,\ldots,N_i \tag{12}$$

The overload distortion is given by [1,2]:

$$D_o = 2\int_{x_{\max}}^{\infty}(x-y_{\max})^2\, p(x)dx\ , \tag{13}$$

where the $y_{\max}$ is determined from the condition:

$$y_{\max} = \frac{\int_{x_{\max}}^{\infty} xp(x)dx}{\int_{x_{\max}}^{\infty} p(x)dx}\ , \tag{14}$$

the property of the constant tail centroids is used in the design process. Finally, combining the set of equations (5), (6), (13) and (14), one can derive the overload distortion of the proposed companding quantizer as in [5]:

$$D_o = \sqrt{\frac{2}{\pi}}\frac{1}{x_{\max}^3}e^{-\frac{x_{\max}^2}{2}}\ . \tag{15}$$

By determining the total distortion $D$ as a sum of the granular distortion $D_g$ (12) and the outer distortion $D_o$ (15), the signal to quantization noise ratio can be determined [1,2]:

$$\mathrm{SQNR} = 10\log\left(\frac{\sigma^2}{D_g + D_o}\right) = 10\log\left(\frac{\sigma^2}{D}\right). \tag{16}$$

## 4. NUMERICAL DETERMINATION OF THE OPTIMAL VALUE OF SEGMENT THRESHOLD

In this section, numerical determination of the optimal value of segment threshold is described. The optimal value of segment threshold $x_i$ is determined in the following:

**Step 1**. Solving the system of equations (3), the expressions for determining coefficents $r_i$, $p_i$, $q_i$, $i = 1,...,L$ are obtained. The optimal value of the support region, $x_{\max}$, is defined by equation (6). For different values of the segment threshold, $\frac{x_{\max}}{2} < x_i < x_{\max}$ ,the values of coefficents $r_i$, $p_i$, $q_i$, $i$=1,...,$L$ are determined.

**Step 2.** Based on the coefficents $r_i$, $p_i$, $q_i$, $i$=1,...,$L$ which are obtained in a method described in step 1, the quadratic spline functions, $h(x)$, defined by equation (1) are obtained.

**Step 3.** On the basis of the quadratic spline functions obtained in step 2, $h(x)$, the proposed companding quantizer as described in section 3 is designed.

**Step 4.** For different values of the segment threshold, $\frac{x_{\max}}{2} < x_i < x_{\max}$, the SQNR value of the proposed companding quantizer which is designed as described in step 3 is determined. The optimal value of the segment treshold is determined depending on the maximal SQNR value of the proposed companding quantizer (see Figure 1).

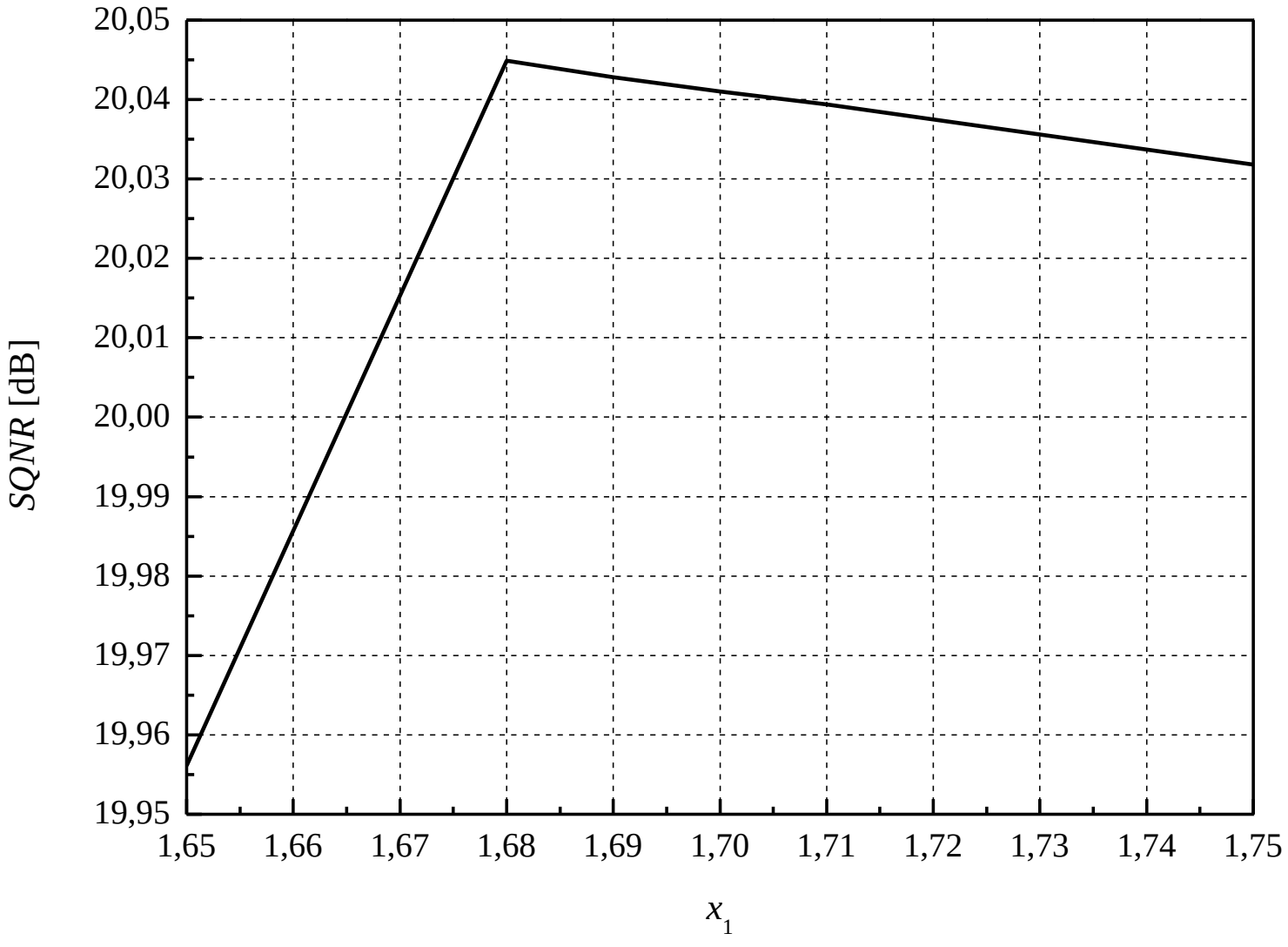

**Fig. 1. Numerical determination the optimal value of the segment threshold $x_1$ for the number of levels $N = 16$.**

## 5. NUMERICAL RESULTS AND CONCLUSION

Numerical results presented in this section are obtained for the case when the number of segments is equal to $2L = 4$ and for number of levels $N = 16$ and $N = 32$. For this number of segments, the approximate quadratic spline function $h(x)$ defined by equation (1) is equal to:

$$h(x) = \begin{cases} r_1 + p_1 x + q_1 x^2, & x \in [0, x_1] \\ r_2 + p_2 x + q_2 x^2, & x \in [x_1, x_2] \end{cases}. \tag{17}$$

Based on equation (17) it can be noticed that for the approximate quadratic spline function $h(x)$ design it is necessary to determine six coefficients, $r_1$, $p_1$, $q_1$, $r_2$, $p_2$, $q_2$. In this paper, these coefficient are determined by minimization mean square error (solving the system of equations defined by equation (3)). The values of coefficents $r_1$, $p_1$, $q_1$, $r_2$, $p_2$, $q_2$ depend on the value of the segment threshold $x_1$. In order for the quadratic spline function to better approximate the optimal compressor function, the value of the segment threshold $x_1$ is numerically determined as described in section 4. Numerical determination of the optimal value of the segment threshold $x_1$ for the number of levels $N = 16$ is presented in figure 1. For the number of segments $2L = 4$ and for number of levels $N = 16$, the optimal value of the segment threshold $x_1$ is equal to $x_1 = 1.68$. In the other case, for the number of segments $2L = 4$ and for number of levels $N = 32$, the optimal value of the segment threshold $x_1$ is equal to $x_1 = 2.25$. For these optimal values of the segment threshold $x_1$, coefficients $r_1$, $p_1$, $q_1$, $r_2$, $p_2$, $q_2$ are determined, the approximate quadratic spline function $h(x)$ is formed and the design of companding quantizer as described in section 3 is done. Table 1 shows the values of SQNR of the proposed companding quantizer which the segment threshold $x_1$ is at the middle of support region, ($SQNR^{EQU}$), the values of SQNR of the proposed companding quantizer which segment threshold $x_1$ numerically determined ($SQNR^{NUM}$), and the values of SQNR of nonlinear optimal companding quantizer [1], ($SQNR^{OPT}$), for the number of levels $N = 16$ and $N = 32$.

**Table 1. The values of $SQNR^{EQU}$, $SQNR^{NUM}$ and $SQNR^{OPT}$ for the number of levels $N$ = 16 and $N$ = 32.**

| N | $SQNR^{EQU}$ [dB] | $SQNR^{NUM}$ [dB] | $SQNR^{OPT}$ [dB] |
|---|---|---|---|
| 16 | 19.69 | 20.04 | 20.22 |
| 32 | 25.80 | 25.99 | 26.01 |

In Figure 2 dependency of $SQNR^{EQU}$, $SQNR^{NUM}$ and $SQNR^{OPT}$ on the number bits per sample is shown.

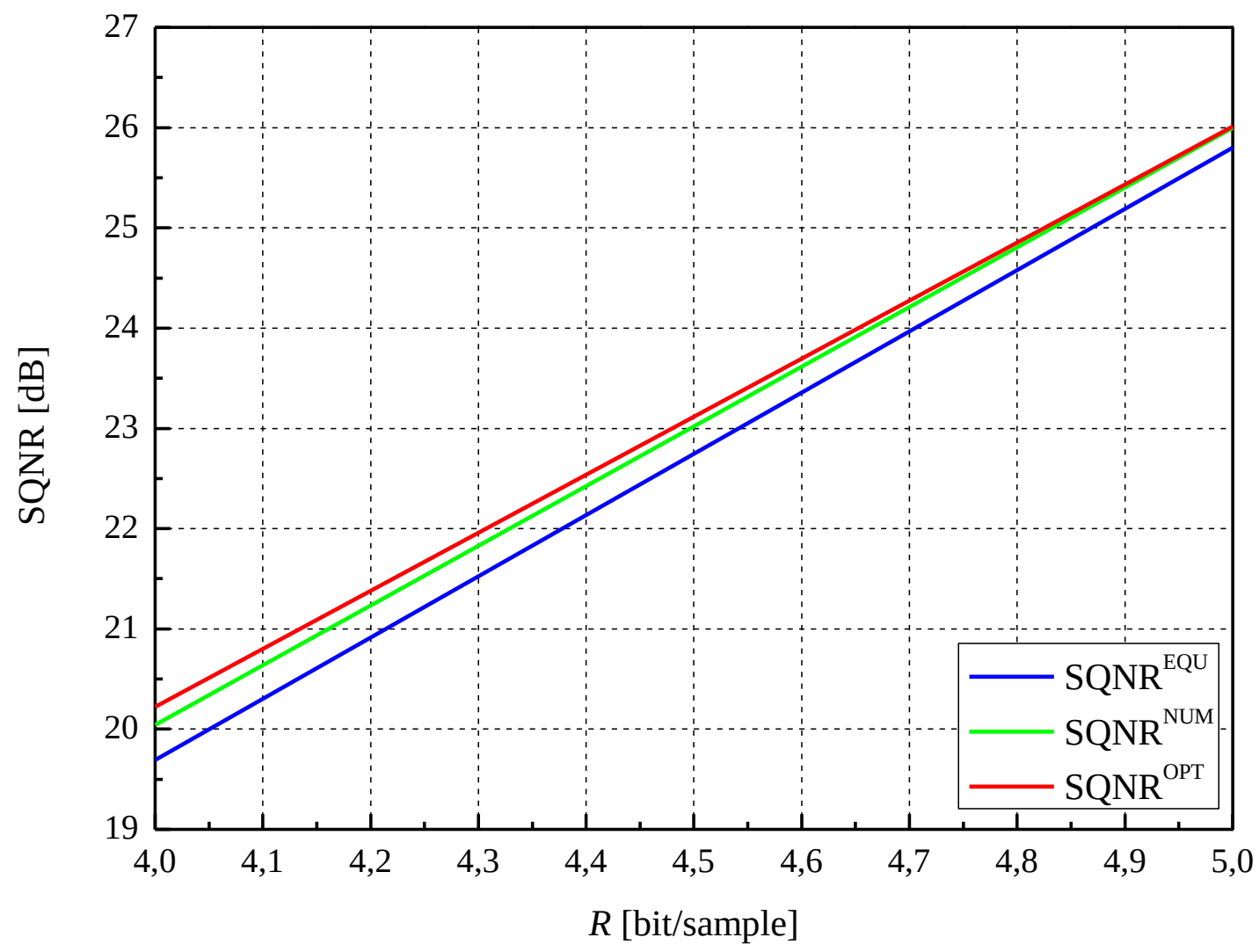

**Fig. 2. Dependency of $SQNR^{EQU}$, $SQNR^{NUM}$ and $SQNR^{OPT}$ on the number bits per sample.**

Analyzing the results shown in Table 1 and Figure 2, one can notice that design of proposed companding quantizer based on approximate quadratic spline functions, with the optimal value of segment threshold $x_1$, achieved higher SQNR than design of quantizer based on approximate quadratic spline functions which coefficients are determined for the case where the segment threshold $x_1$ is at the middle of the support region. Also, one can conclude that design of quantizer based on approximate quadratic spline functions, with the optimal value of segment threshold $x_1$, achieved SQNR very close to that of nonlinear optimal companding quantizer. Since the complexity of the proposed companding quantizer is very low, and an achieved SQNR very close to that of nonlinear optimal companding quantizer, it can be concluded that the proposed companding quantizer presents a very efficient solution.

## ACKNOWLEDGEMENTS

This work is partially supported by Serbian Ministry of Education and Science through Mathematical Institute of Serbian Academy of Sciences and Arts (Project III44006) and by Serbian Ministry of Education and Science (Project TR32035).